# Cepstrum-Based Texture Features for Melanoma Detection

Keith Miller, Tristan Crawford, Jason Hagerty, PhD, William Stoecker, MD, Ronald J. Stanley, PhD

*Abstract*— **This paper introduces a set of cepstrum-based texture features for melanoma classification and validates their performance on dermoscopic images from the ISIC 2019 dataset. We propose applying gray-level co-occurrence matrix (GLCM) statistics to 2D cepstral representations—a novel approach in image analysis. Combined with established handcrafted lesion descriptors, these features were evaluated using XGBoost models. Incorporating select cepstral features improved the area under the receiver operating characteristic curve, accuracy, and F1 score for binary melanoma vs. nevus classification. Results suggest that cepstral GLCM features offer complementary discriminatory information for melanoma detection.**

## I. INTRODUCTION

The cepstrum was initially introduced for one-dimensional signal processing by Bogert et al. [1]. Later, Oppenheim et al. [2] applied cepstral analysis techniques to image processing. Cepstral analysis has since been explored in various contexts, including anomaly detection and pattern classification. The cepstrum is derived from the inverse Fourier transform of the log magnitude of the Fourier transform of a signal (Equ 1).

$$C_t = Re\{F^{-1}\{\log|F\{x(t)\}|\}\}$$

Equation 1: real cepstrum of x(t)

Bogert et. al. showed that for simple echo signals (represented by the sum of a signal and the same signal scaled by a constant α and time delayed by τ seconds, see equ 2), time delayed terms will become sinusoidal signals with frequency τ, in the frequency domain, after a logarithm is applied to the Fourier transform. If an inverse transform is then applied, the sinusoidal frequency domain signals resulting from the time shift become impulses at τ.

$$y(t) = x(t) + \alpha x(t - \tau)$$

Equation 2: simple echo signal

Because of this property of converting time delays to impulses, the cepstrum captures repeating spatial/frequency structures, such as echoes, rings, or grids, which manifest as peaks or periodic patterns in the cepstral domain (Fig. 1). Cepstral analysis offers a powerful method for extracting texture information from images, capturing periodic patterns and harmonics that are often difficult to characterize in the spatial domain (Fig. 2). However, cepstral methods remain underutilized in medical image analysis, particularly in dermatology. In dermoscopic images, cepstral features may correspond to repeating physical structures such as pigment networks, globules, or streaks.

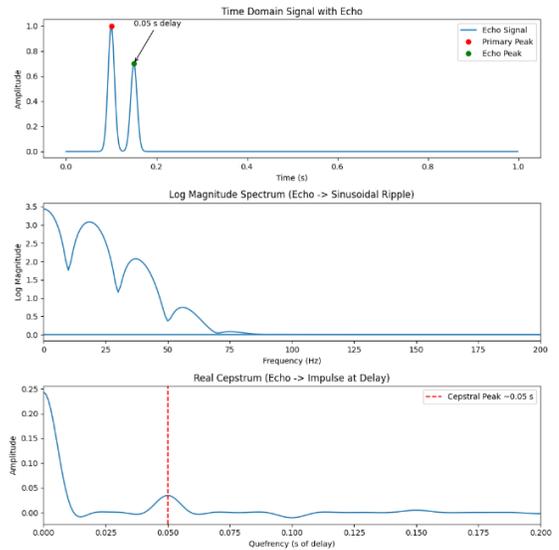

Figure 1. Example cepstrum of a simple echo.

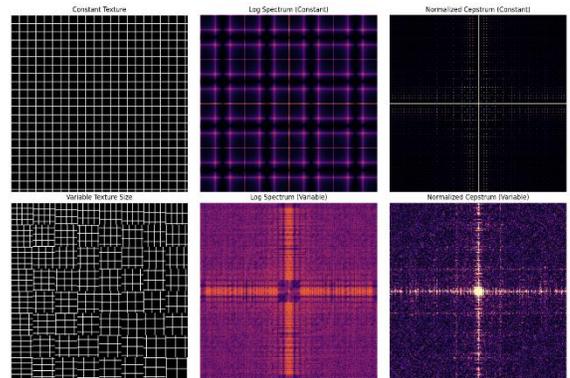

Figure 2. Example change in cepstra due to textural variance

Recent research has used the cepstrum as input to convolutional neural network classifiers [3] with promising results, but interpretable cepstrum-derived features have not been explored for skin lesion classification

In this paper, we introduce a novel approach to image classification that utilizes gray-level co-occurrence matrix (GLCM) analysis on the 2D cepstrum of dermoscopic images. While Haralick texture features derived from GLCMs have a long history in texture analysis [4], their application to cepstral analysis is novel. This fusion of cepstral and GLCM features captures local spatial dependencies, textural transitions, and global periodicity across scales, thereby increasing the information content of the feature set alongside texture features for lesion characterization.

To evaluate the effectiveness of cepstrum-based texture descriptors, we conducted an ablation study using handcrafted features that represent both clinically and statistically observed melanoma traits. These include median-split color metrics, atypical pigment network (APN), and pink shade features, drawn from established work on mixed-domain feature fusion for skin disease classification [5].

We train XGBoost models on both individual and combined feature groups extracted from the ISIC 2019 challenge dataset [6][7][8], focusing on the binary task of melanoma versus nevus classification. Results show that cepstral texture features offer a measurable improvement in classification performance when combined with conventional handcrafted descriptors, highlighting their potential as a complementary modality for skin lesion analysis.

## II. CEPSTRAL FEATURES

We propose a 35-element cepstrum-derived feature set that captures statistical, structural, and directional properties of dermoscopic lesions in the frequency domain. These features are extracted from the real 2D cepstrum of lesion image channel pairs across multiple color spaces, enabling robust analysis of periodic structures and fine texture not readily discernible in the spatial domain. By applying gray-level co-occurrence matrix (GLCM) statistics to the cepstrum, we derive descriptors that measure texture anisotropy and harmonic regularity—complementing conventional Haralick features extracted from pixel intensities, across all structural scales, eliminating the need to tune GLCM pixel distance to any pixel distance other than one, which is required for most analyses using GLCM features. Additional non-GLCM cepstral features, including global statistics (mean, variance, skew, kurtosis), cepstral entropy, and radial profile metrics (peak value and area under the curve (AUC) for the receiver operating characteristic (ROC), and contribute interpretable summaries of frequency structure within the lesion mask.

### A. Preprocessing

Each dermoscopic image was converted into multiple color spaces: RGB, Lab, HSV, and YCrCb were used for this experiment. The individual channels of each color space were processed independently. Representing the image in multiple color spaces allowed for increased insight into model and feature performance. We have also seen that including features from multiple color spaces resulted in increased classifier performance.

For each lesion, we generated a binary lesion mask using the algorithm presented by Norsang Lama [9], which segments the lesion from the surrounding skin (Fig. 3). These masks were used to isolate the lesion region prior to cepstral analysis (Fig. 3), ensuring that the cepstrum predominantly captured texture patterns from within the lesion itself. This preprocessing step significantly reduces background interference and enhances the reliability of extracted cepstral features. For each channel of each space, a normalized unsigned integer representation of the cepstrum is also generated for GLCM feature extraction.

### B. Non-GLCM Cepstral Features

For each lesion spectrum, the following general metrics are extracted:
- Mean
- Standard Deviation
- Skew
- Kurtosis
- Entropy
- Radial Peak Value
- Radial Area-Under-the-Curve

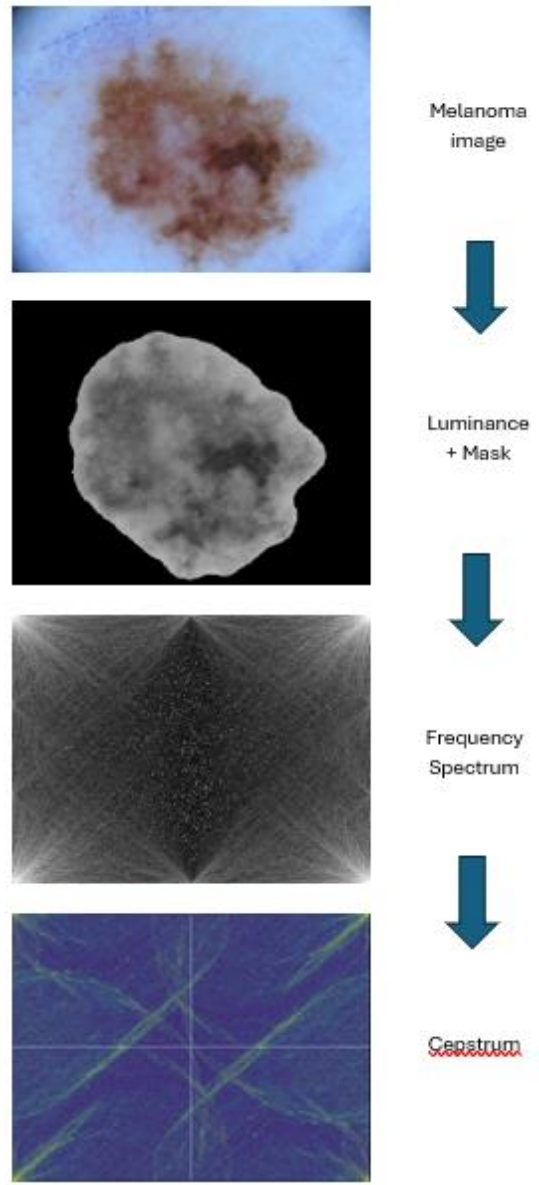

Figure 3. Preprocessing pipeline for extracting the 2D cepstrum. masks are applied per channel, and cepstra are extracted for each channel of the image.

## C. GLCM Cepstral Features

To further exploit spatial-frequency structure over the lesions, we computed Gray-Level Co-occurrence Matrix (GLCM) statistics over the 2D real cepstrum. While GLCM features have been extensively used in texture analysis of spatial-domain images, their application to cepstral images is novel and motivated by the distinct structures observable in the cepstrum, such as streaks and rings.

The GLCM is computed for pixel pairs in four directions (0°, 45°, 90°, and 135°, see figure 4). From these, we calculate the thirteen stable statistical texture features described by Haralick et al. [4], including:

*Contrast* – Measures local intensity variations within the cepstrum.

*Correlation* – Captures linear dependencies of pixel pairs.

*Energy (Angular Second Moment)* – Reflects cepstral homogeneity.

*Entropy* – Quantifies disorder or complexity in cepstral texture.

*Homogeneity (Inverse Difference Moment)* – Sensitive to the proximity of similar cepstral magnitudes.

*IMC1 / IMC2* – Information-theoretic metrics capturing complex dependencies.

*Sum Entropy / Difference Entropy* – Reflect high-order textural variation.

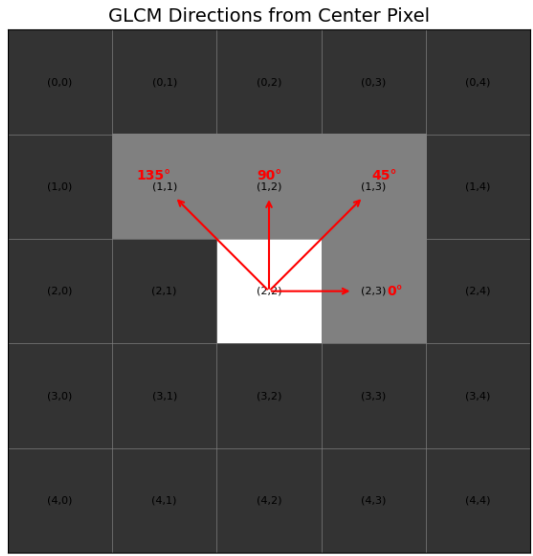

Figure 4. Example GLCM pixel pair directions.

We also calculate the matrix Trace for each directional GLCM [10]. Pixel pair directions are shown in Fig. 4 This yields fourteen features per GLCM direction. The mean overall directions and absolute maximum are retrieved per feature, and the mean and max/mean quotient is saved. The max/mean quotient serves as an anisotropy metric for the feature; we will refer to this as the directionality of the feature. The Haralick features are calculated using the mahotas library [11].

These twenty-eight features are computed per channel and per color space (BGR, Lab, HSV, YCrCb), resulting in a rich and interpretable set of direction-sensitive texture descriptors.

These features provide a means of capturing *anisotropic or repetitive structures* often observed in malignant lesions.

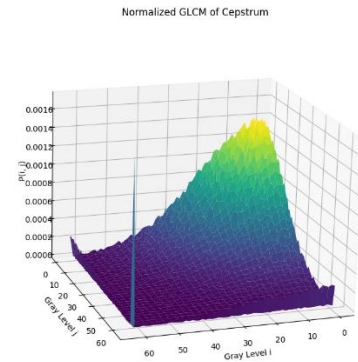

Figure 5. Plot of GLCM of the cepstrum shown in Image 3. The distribution is higher for lower value pixel pairs ( 1-1, 2-2, etc.). This result is consistent with the nature of the cepstrum image: dark and homogeneous.

## D. Statistical Analysis of Features

To evaluate the discriminatory potential of our proposed cepstral features, we conducted a statistical analysis against the binary melanoma/nevus labels. Two complementary metrics were used:

| Feature | Correlation | MI |
|---|---|---|
| RGB_C1_std | -0.206628 | 0.028711 |
| RGB_C0_std | -0.206486 | 0.026291 |
| RGB_C1_radial_peak_val | -0.202736 | 0.0252 |
| RGB_C1_radial_AUC | -0.201938 | 0.023011 |
| YCrCb_C0_std | -0.197316 | 0.025941 |
| Lab_C0_std | -0.197165 | 0.015626 |
| Lab_C2_Har_Cep_sum_entropy_Dir | 0.193214 | 0.015082 |
| Lab_C1_Har_Cep_sum_entropy_Dir | 0.193124 | 0.014441 |
| YCrCb_C2_Har_Cep_sum_entropy_Dir | 0.192763 | 0.010814 |
| YCrCb_C2_std | -0.192615 | 0.029211 |
| YCrCb_C2_cepstral_entropy | 0.192228 | 0.029043 |
| YCrCb_C1_Har_Cep_sum_entropy_Dir | 0.191611 | 0.015355 |
| YCrCb_C1_cepstral_entropy | 0.190269 | 0.022948 |
| YCrCb_C1_Har_Cep_contrast | 0.189437 | 0.02101 |
| YCrCb_C2_Har_Cep_contrast | 0.186935 | 0.020534 |
| Lab_C2_cepstral_entropy | 0.186369 | 0.026341 |
| Lab_C1_cepstral_entropy | 0.184459 | 0.024892 |
| Lab_C1_Har_Cep_contrast | 0.183468 | 0.020109 |
| Lab_C2_Har_Cep_contrast | 0.183404 | 0.015772 |
| RGB_C0_radial_peak_val | -0.183114 | 0.01735 |

Table 1. Top 20 Cepstral features by absolute Pearson correlation

1. *Pearson correlation* – captures the linear association between individual features and class labels.

2. *Mutual information (MI)* – quantifies the nonlinear dependence between feature distributions and the target classes, reflecting their potential contribution to classification.

Table 1 lists the top 20 features ranked by absolute Pearson correlation with their respective MI. Negative correlation values indicate that higher feature magnitudes are associated with the nevus class, while positive correlations reflect melanoma-biased patterns. Notably, features such as RGB_C1_std, RGB_C0_std, and radial peak/AUC metrics from the RGB channels showed the strongest negative correlations with melanoma, suggesting that lesions with lower within-channel variance or weaker radial cepstral responses are more likely benign. This may be due to "crisp" non-varying textures having more pronounced impulses in the cepstral domain and a lower mean value, which will result in a higher standard deviation over the cepstrum.

Conversely, cepstral entropy and GLCM-derived Haralick features in Lab and YCrCb color spaces exhibited positive correlations with melanoma, consistent with the expectation that malignant lesions present more heterogeneous and multi-scale texture patterns in the cepstrum. Four of the features (Lab_C2_Har_Cep_sum_entropy_Dir, Lab_C1_Har_Cep_sum_entropy_Dir, YCrCb_C2_Har_Cep_sum_entropy_Dir, YCrCb_C1_Har_Cep_sum_entropy_Dir) are novel Haralick on cepstrum directionality features, demonstrating a correlation between anisotropic texture variation and the malignant class. Figure 6 (Top 20 Features by Mutual Information) visualizes the top 20 features ranked by mutual information. We observe that mutual information and correlation trends do not necessarily align, with MI, further emphasizing nonlinear contributions. Several novel *Haralick-on-cepstrum* features appear among the top-ranked features, including *IMC1* of the cepstrum of the HSV Saturation and Value channels, as well as *sum/difference entropy metrics*, validating their relevance in capturing anisotropic and scale-sensitive lesion structures.

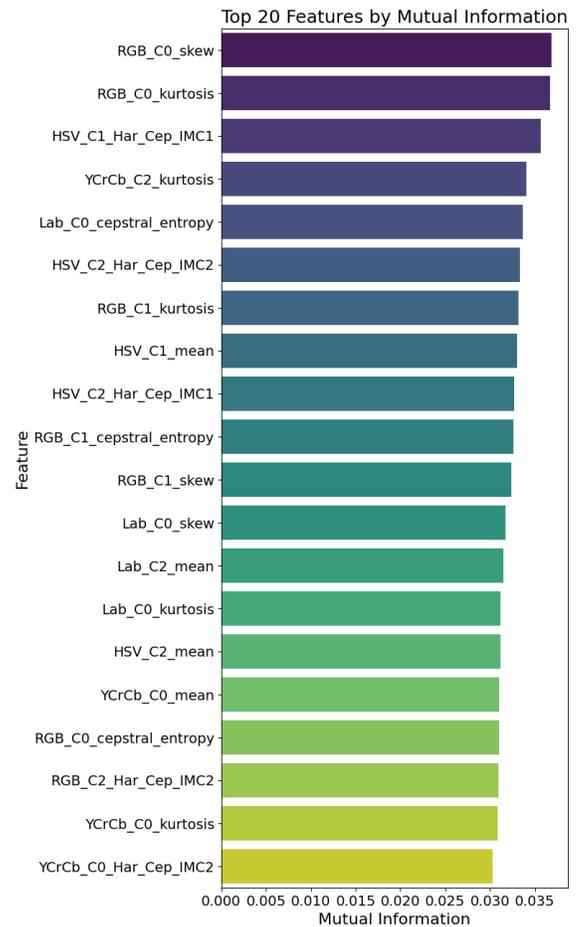

Figure 6. Highest cepstral feature MI scores. All twenty highest scoring features have an MI above 0.03.

The concentration of high-MI features among YCrCb and Lab channels suggests that these color spaces, which better decouple luminance and chromaticity, provide richer cepstral descriptors for skin lesions.

### III. EXISTING HANDCRAFTED FEATURES

In addition to the novel cepstral features, we incorporated several handcrafted feature groups previously developed for dermoscopic image analysis [12].

**Median-Split Color Metrics**: These features describe the distribution and shape of color regions resulting from lesion quantization using the median-split algorithm [13]. Overall, this method attempts to capture color asymmetry.

**Atypical Pigment Network (APN) Features** [14]: APNs were segmented using a threshold based on the variance of the red channel from the RGB lesion image. Once localized, features were developed to quantify network patterns via measurements of describing color, density, and symmetry of the networks throughout the image. The likelihood of a melanoma diagnosis increased with an increased presence of APN.

**Pink Region Features** [15]: A soft color thresholding method is used to isolate pink or reddish hues. Variance, area fraction, and location-based statistics are used to detect inflammation, blood vessels, and regression structures within lesions, which are frequent in malignant cases.

**Saliency Features**: These metrics are derived from the binarization of the Laplacian-of-Gaussian (LoG) saliency maps. Using the binary image, object count, spacing, and shape irregularity are quantified. Highly salient but irregular structures, such as dots or streaks, are informative indicators of melanoma.

**Telangiectasia Detection**: Based on the method in [16], these features identify red structures corresponding to small blood vessels (telangiectasia). These are often seen in basal cell carcinoma but are also relevant to melanoma classification.

**Semitranslucent Area Features**: Attempts to quantify lesion area with a "smooth, jelly-like area with varied, near-skin-tone color, can indicate a diagnosis of basal cell carcinoma" [17].

Each of these feature families captures a distinct visual cue used by dermatologists, and together they form a comprehensive baseline for evaluating new feature types such as cepstrum-derived descriptors.

## IV. EXPERIMENTAL SETUP AND RESULTS

For this study, we used dermoscopic images from the *ISIC 2019 Challenge Dataset*, focusing on the binary classification task of *melanoma vs. nevus*. To improve generalizability and reduce bias due to image duplication, we applied the following filtering and stratification steps:

*Class Selection*: Only images labeled as *melanoma* or *nevus* were retained for this binary classification task.

*Duplicate Handling*: Many dermoscopic images in the ISIC dataset appear multiple times under different image IDs. We applied a filtering process to identify and isolate *unique lesion images* by matching lesion IDs. A subset of *unique images* were reserved for testing. *Duplicate images* were used exclusively in training to avoid data leakage.

*Final Dataset Split*: The dataset was divided into training and test subsets using an 80:20 split. The test set consisted only of unique lesions that were not present in the training set.

This approach ensures that model evaluation is conducted on images unseen during training.

To identify the highly discriminative features from a large set of handcrafted and cepstral descriptors, we applied a *greedy forward selection algorithm* using ROC AUC as the evaluation metric.

| Feature Set | Accuracy (Base) | Accuracy (+Cepstrum) | F1 Score (Base) | F1 Score (+Cepstrum) | ROC AUC (Base) | ROC AUC (+Cepstrum) |
|---|---|---|---|---|---|---|
| **APN** | 0.9146 | **0.9376** | 0.4936 | **0.5500** | 0.9165 | **0.9251** |
| **Pink** | 0.9216 | **0.9298** | 0.5249 | **0.5345** | 0.9170 | **0.9289** |
| **Median** | 0.9367 | **0.9406** | **0.5899** | 0.5886 | **0.9488** | 0.9435 |
| **Salient** | 0.8631 | **0.9181** | 0.3804 | **0.4676** | 0.8528 | **0.9061** |
| **Telangiectasia** | 0.7860 | **0.9155** | 0.2492 | **0.4144** | 0.6854 | **0.8946** |
| **Semitranslucent** | 0.9168 | **0.9341** | 0.5077 | **0.5503** | 0.9178 | **0.9228** |

Table 2. Feature set performance with and without cepstral features

The greedy feature selection procedure works as follows:

1. **Initialization**: Start with an empty set of selected features.
2. **Iteration**:
   a. At each step, evaluate all remaining features by temporarily adding them to the current selection set.
   b. Train an **XGBoost classifier** on the selected features and measure **ROC AUC** on a validation set.
   c. Select the feature that yields the greatest AUC improvement and add it to the selected set.
3. **Stopping Criterion**: Stop once a predefined number of features is selected (e.g., 50 or 100).

This method allows the identification of feature combinations that are not only individually informative but also synergistic in improving model performance. It also helped identify **redundant or non-contributory features**, especially among those with high intercorrelation.

### A. Multi-Feature-Set Performance

We trained XGBoost classifiers using different feature subsets to evaluate the effectiveness of cepstral descriptors. Results for each feature set are summarized in Table 2. In addition, cepstral features alone achieved an accuracy of 0.9103, F1 score of 0.3591, and ROC AUC of 0.8811.

These results demonstrate that cepstral features consistently improve diagnostic performance when combined with existing handcrafted descriptors. The most notable improvements occur with weaker feature groups like telangiectasia and salient features, indicating the strength of cepstral features in capturing complementary image properties.

However, when all feature sets were combined, model performance did not improve with the addition of cepstral features. This suggests that careful feature selection may be necessary to avoid redundancy and overfitting when merging diverse handcrafted and cepstral features.

Using the previously outlined greedy feature selection process significantly improved the XGBoost model's ROC AUC, with the highest-performing model achieving an AUC of 0.9697, utilizing fewer than 100 total features selected from all feature sets. Higher performing feature sets were achieved with the inclusion of cepstral features (see Figs. 7, 8) without increasing the size of the resulting feature set. This suggests our novel features are contributing new information, allowing the XGBoost model to segment the feature space more efficiently.

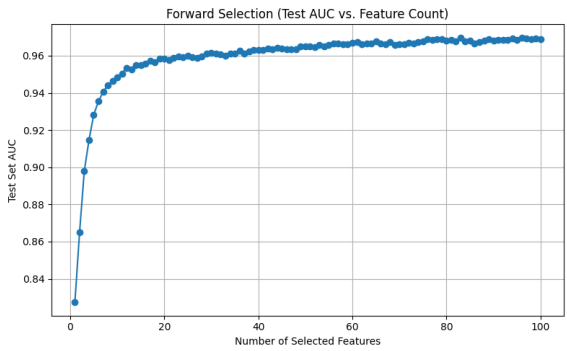

Figure 7. Graph of ROC AUC per iteration of the greedy feature selection algorithm applied to all features (cepstral and non-cepstral). Model performance plateaus with an AUC of 0.969; additional features lead to model degradation.

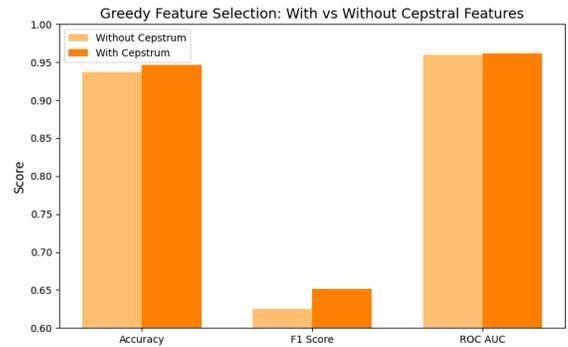

Figure 8. Greedy selected feature model performance with and without cepstral features; 100 features were selected for each model. cepstral features contribute considerable improvements to F1 and Accuracy, with a slight improvement to ROC AUC

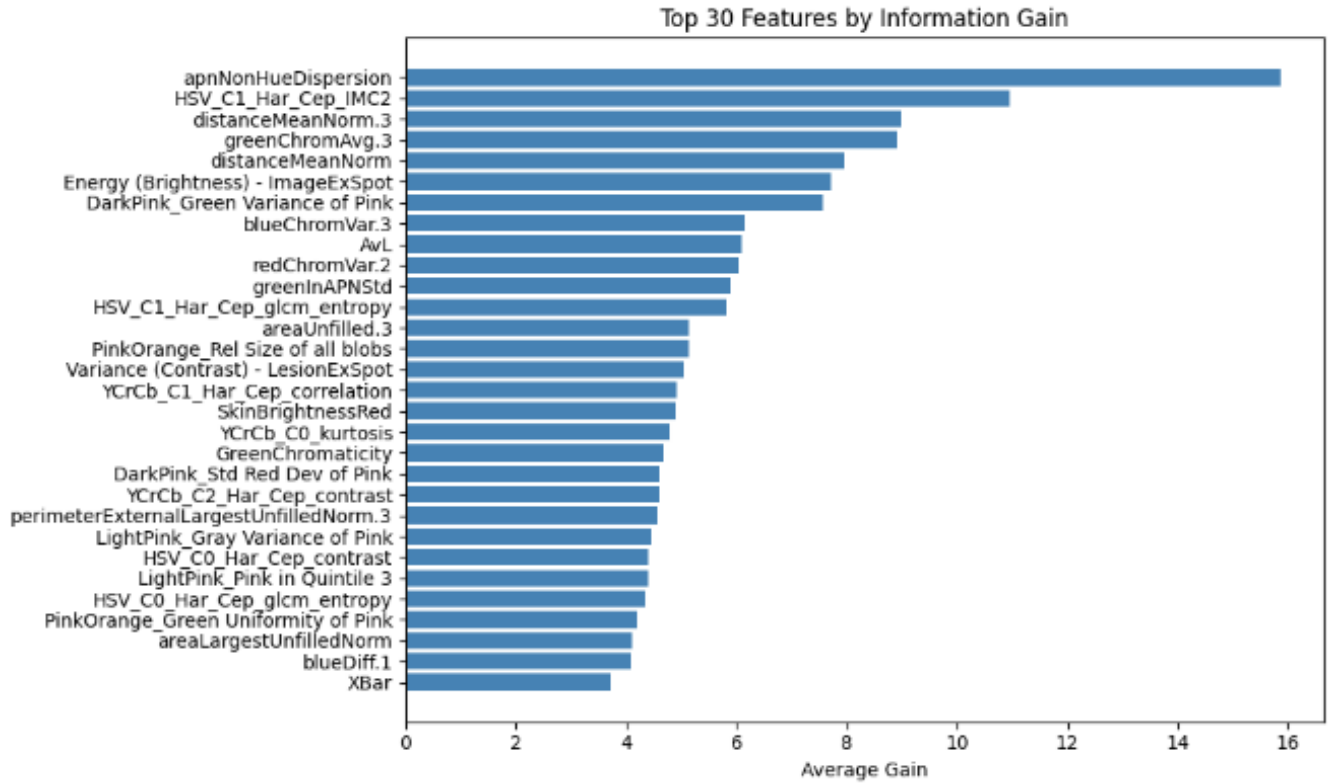

Figure 9. Average information gain for the highest performing 100-feature model. 7 of the top 30 features are from the novel cepstral feature set.

## V. Conclusion

In this work, we introduce a novel set of interpretable cepstrum-based texture features for melanoma classification, encompassing both statistical and spatially sensitive descriptors. Our approach uniquely combines 2D cepstral analysis with Haralick-style GLCM statistics, revealing anisotropic and frequency-domain patterns that correlate with malignancy.

Through careful preprocessing, feature engineering, and classifier benchmarking, we demonstrated that selected cepstral features consistently improved performance across multiple handcrafted feature sets. highlighting their diagnostic utility.

While the addition of cepstral features often yielded modest gains in ROC AUC, it consistently improved F1 scores — suggesting their utility in reducing false negatives. Cepstral features regularly had large average information gain in their respective XGBoost models. These findings support the use of cepstrum-based texture features as a complementary tool in dermoscopic analysis, and open avenues for further integration with deep learning pipelines or real-time screening systems.

Future work includes evaluating cepstral features in multiclass classification tasks (e.g., distinguishing melanoma from other malignancies), exploring alternative frequency-domain representations such as the 2D 'Short Time' cepstral transform, and integrating these descriptors into lightweight deep learning architectures for real-time screening. We also plan to study the clinical interpretability of high-ranking cepstral features in collaboration with dermatologists, as well as apply these novel features to other medical imaging tasks.